\documentclass[reprint,amsmath,amssymb,aps,prl]{revtex4-1}

\usepackage{graphicx}
\usepackage{dcolumn}
\usepackage{bm}
\def\ph2{{\it p}-H$_2$}
\def\od2{{\it o}-D$_2$}

\begin{document}
\title{Superfluid response of parahydrogen clusters in superfluid $^4$He}

\author{Massimo Boninsegni}
\affiliation{%
Department of Physics, University of Alberta, Edmonton, Alberta, T6G 2E1, Canada}%

\date{\today}

\begin{abstract}
First principle computer simulations yield evidence of a {\em finite} superfluid response at low temperature
of a parahydrogen cluster of 15
molecules in bulk superfluid $^4$He.  The superfluid fraction  is worth  $\sim 44$\% at $T=0.25$ K, growing to about $\sim 66\%$ at  $T=0.15$ K, i.e., it is substantially reduced compared to that of the same cluster
{\em in vacuo}, due to higher molecular localization. The implications of these findings on the interpretation of experiments with linear molecules embedded in parahydrogen clusters immersed in  superfluid helium  are discussed. 

\end{abstract}
\maketitle

Almost half a century since the original prediction of Ginzburg and Sobyanin \cite{gs} of a possible superfluid transition at low temperature of a fluid of parahydrogen (\ph2) molecules, its observation has eluded even the ablest experimenters and/or cleverest approaches. Meanwhile, there is now a wealth of robust theoretical predictions, based on state-of-the-art many-body techniques and realistic intermolecular potentials, pointing to the following: \\ \indent
{\em a}) molecular hydrogen has a strong tendency to crystallize at temperatures well above those at which Bose condensation and superfluidity (SF) should occur, even in reduced dimensions \cite{boninsegni04,boninsegni13}, disorder \cite{boninsegni05,turnbull,boninsegni16} and confinement \cite{screw,omiyinka}  
\\ \indent
{\em b}) even if a metastable fluid phase of \ph2 could be stabilized well below its 13.8 K freezing temperature \cite{note}, it may only  turn superfluid at temperatures at least two orders of magnitude lower than that (few K) predicted in the original 1973 work \cite{boninsegni18}. \\ \indent
To date, the only quantitative prediction of superfluid behavior of \ph2 has been made not for the bulk phase, but for small clusters (few tens of molecules), at temperatures of the order of 1 K \cite{sindzingre,fabio,fabio2,fabio3}. 
Experimentally, the superfluid behavior of a quantum cluster can be inferred from the free rotation of a linear molecule embedded in it, as shown by high-resolution microwave or infrared spectroscopy \cite{historic}. This methodology has allowed for the remarkable observation of superfluidity in $^4$He clusters of just a few atoms \cite{tang}, and has also yielded some evidence of possible superfluid behavior of \ph2 clusters comprising around fifteen molecules
 \cite{grebenev,li,raston}, in some cases backed by theoretical results \cite{li,kwon}. In some of these experiments \cite{grebenev,grebenev2,grebenev3,grebenev4,scassa}, the \ph2 clusters are immersed in relatively large superfluid $^4$He nanodroplets (comprising several hundreds to several thousand atoms), rendering this a potentially unique example of a mixture of two  superfluid Bose components.
 \\ \indent 
The interpretation in terms of superfluidity of the \ph2 clusters has been questioned, however, in the case of experiments in which the cluster is embedded in a $^4$He matrix; in particular, recent simulation studies of small, mixed \ph2/$^4$He clusters with an embedded CO$_2$ dopant, suggest that doped \ph2 clusters may form a non-superfluid core in $^4$He droplets \cite{mavaff}, and that $^4$He acts to suppress \ph2 superfluidity in small, mixed clusters, a conclusion previously reached by others \cite{gordillo}.
\\ \indent
The question of the possible superfluidity of \ph2 clusters in superfluid $^4$He remains open, however, as there are at least two aspects that warrant further investigation. The first is that the simulations of Ref. \onlinecite{mavaff} were carried out at temperature $T=0.5$ K, considerably higher than that ($T=0.15$ K) at which the onset of the superfluid response is reported in Ref. \onlinecite{grebenev}. Second, the behavior of a small mixed cluster with a comparable number (of the order of ten) of $^4$He atoms and \ph2 molecules might be qualitatively different from that of the same number of \ph2 molecules in a $^4$He matrix of size such that its physical properties approach that of bulk $^4$He. Indeed, to our knowledge there are presently no theoretical predictions regarding structural and superfluid properties of \ph2 clusters in superfluid $^4$He. 
\\ \indent
In this paper, we present results of first principle computer simulations of a (\ph2)$_{15}$ cluster in bulk superfluid $^4$He, at temperature as low as $T=0.15$ K. The purpose of this study is to gain further insight into this intriguing system, and specifically on the effect of the $^4$He matrix on the superfluid properties of (\ph2)$_{15}$, which {\em in vacuo} is predicted to  be superfluid at relatively high temperature (close to 70\% at $T=2$ K, and $\sim$ 100\% at $T=1$ K \cite{notex}). Obviously, a second important goal is to weigh in on the existing controversy regarding the experiment of Ref. \onlinecite{grebenev}, by providing an upper bound on the superfluid response of a pristine (\ph2)$_{15}$ cluster in $^4$He.
\\ \indent
We find here that the structure of the (\ph2)$_{15}$ cluster in superfluid helium is fairly similar to that of a free (\ph2)$_{15}$ cluster (i.e., {\em in vacuo}), the most significant difference being that \ph2 molecules are more localized, especially in the inner shell, and as a result the superfluid response of the cluster is suppressed. It is not completely eliminated, though; for, at temperatures of the order of 0.3 K quantum-mechanical exchanges of \ph2 molecules, virtually absent at $T=0.5$ K begin to occur, and the superfluid signal picks up, growing monotonically as $T\to 0$. At $T=0.15$ K, the superfluid fraction  of the cluster is close to 65\%. This behavior is broadly consistent with the observations of Ref. \onlinecite{grebenev}.
\\ \indent
It is important to note that the structure of the cluster remains essentially unchanged in the low temperature limit, even as exchanges become frequent and a finite superfluid response emerges. In other words, the system is not observed to undergo ``quantum melting" , i.e., going from solid to liquidlike , as observed in simulations of some free clusters \cite{fabio}. Rather, the physical behavior observed here in simulation, with simultaneous localization and superfluidity, is reminiscent of that of the free (\ph2)$_{26}$ cluster, for which the denomination ``supersolid" was proposed \cite{fabiosupersolido}. 
\\ \indent
We now discuss our study in detail. We considered an ensemble of ${\cal N}=\sum_\alpha N_\alpha$ pointlike particles, $\alpha$=$^4$He, \ph2; all particles have spin zero, i.e., all components obey Bose statistics. The system is enclosed in a cubic cell, with periodic boundary conditions in the three directions (as we are simulating a \ph2 cluster inside {\em bulk} $^4$He). The volume of the cell $\Omega$ is adjusted to make the total density ${\cal N}/\Omega$ equal to the equilibrium density of $^4$He at $T=0$, i.e., 0.02183 \AA$^{-3}$.
The quantum-mechanical many-body Hamiltonian reads as follows:
\begin{equation}\label{u}
\hat H = -\sum_{i\alpha} \lambda_\alpha\nabla^2_{i\alpha}+ \frac{1}{2}\sum_{\alpha,\beta}\sum_{ i,j}v_{\alpha\beta}({\bf r}_{i\alpha},{\bf r}_{j\beta}) 
\end{equation}
where $\lambda_\alpha$=6.060 (12.031) K\AA$^{2}$ if $\alpha = $ $^4$He (\ph2), ${\bf r}_{i\alpha}$ is the position of the $i$th particle of component $\alpha$, $v_{\alpha\beta}$ is the Aziz \cite{aziz} (Silvera-Goldman \cite{SG}) pair potential to describe the interaction between two $^4$He atoms (two \ph2 molecules), and we make use of a  potential proposed by Barnett and Whaley \cite{barnett} to represent the interaction between a $^4$He atom and a \ph2 molecule. Although in principle different choices for {\em all} the pair potentials are possible, that made here is consistent with previous studies (see, for instance, Ref. \onlinecite{gordillo}). Moreover, experience accumulated over decades of theoretical studies of both helium and parahydrogen suggests that the most important physical aspects are independent of the detailed form of the potentials utilized.
\\ \indent
We performed first principles QMC simulations of  the system described by Eq.  (\ref{u}), based on the continuous-space Worm Algorithm (WA) \cite{worm,worm2}.  Since this technique is by now fairly well-established, and extensively described in the literature, we shall not review it here. A canonical variant of the algorithm was utilized, in which the numbers $N_\alpha$ of particles for each components are held fixed \cite{fabio,fabio2}.
Details of the simulation are  standard; for instance,  the short-time approximation to the imaginary-time propagator used here is accurate to fourth order in the time step $\tau$ (see, for instance, Ref. \onlinecite{jltp}). The results shown here are obtained with a value of $\tau=(1/320)$ K$^{-1}$, which has been empirically found to yield results indistinguishable from those extrapolated to the $\tau=0$ limit, within our statistical uncertainties. The total number of particles in a typical simulation is ${\cal N}=256$, the number of \ph2 molecules being, of course, equal to 15.
\\ \indent
Besides energetic and structural properties of the (\ph2)$_{15}$ cluster, of special interest here is the superfluid response of both the cluster as well as of the $^4$He matrix. The superfluid fraction of bulk $^4$He is computed through the use of the standard ``winding number" estimator \cite{pollock}. 
On the other hand, superfluidity of a cluster (i.e., a finite system) is defined through its linear response  to an externally imposed rotation around an axis going through the center of mass \cite{leggett}. Specifically, the superfluid fraction of the cluster is defined as $\rho_S=1-(I/I_{cl})$, where $I$ is the moment of inertia of the cluster with respect of the rotation axis, and $I_{cl}$ is its corresponding classical value. Within the path integral formulation of equilibrium statistical mechanics \cite{feynman}, this definition leads to a relatively simple estimator of $\rho_S$ (known as ``area"), which is proportional to the square of the area swept by the many-particle paths in imaginary time \cite{sindzingre89}. Virtually {\em all} quantitative calculations of superfluidity for quantum clusters (mostly helium and parahydrogen) reported in the literature, have been carried out by utilizing the area estimator, in the context of numerical (Monte Carlo) evaluation of path integrals. We make use of the ``area" estimator in this work as well.
\begin{figure}[ht]
\centering
\includegraphics[width=\linewidth]{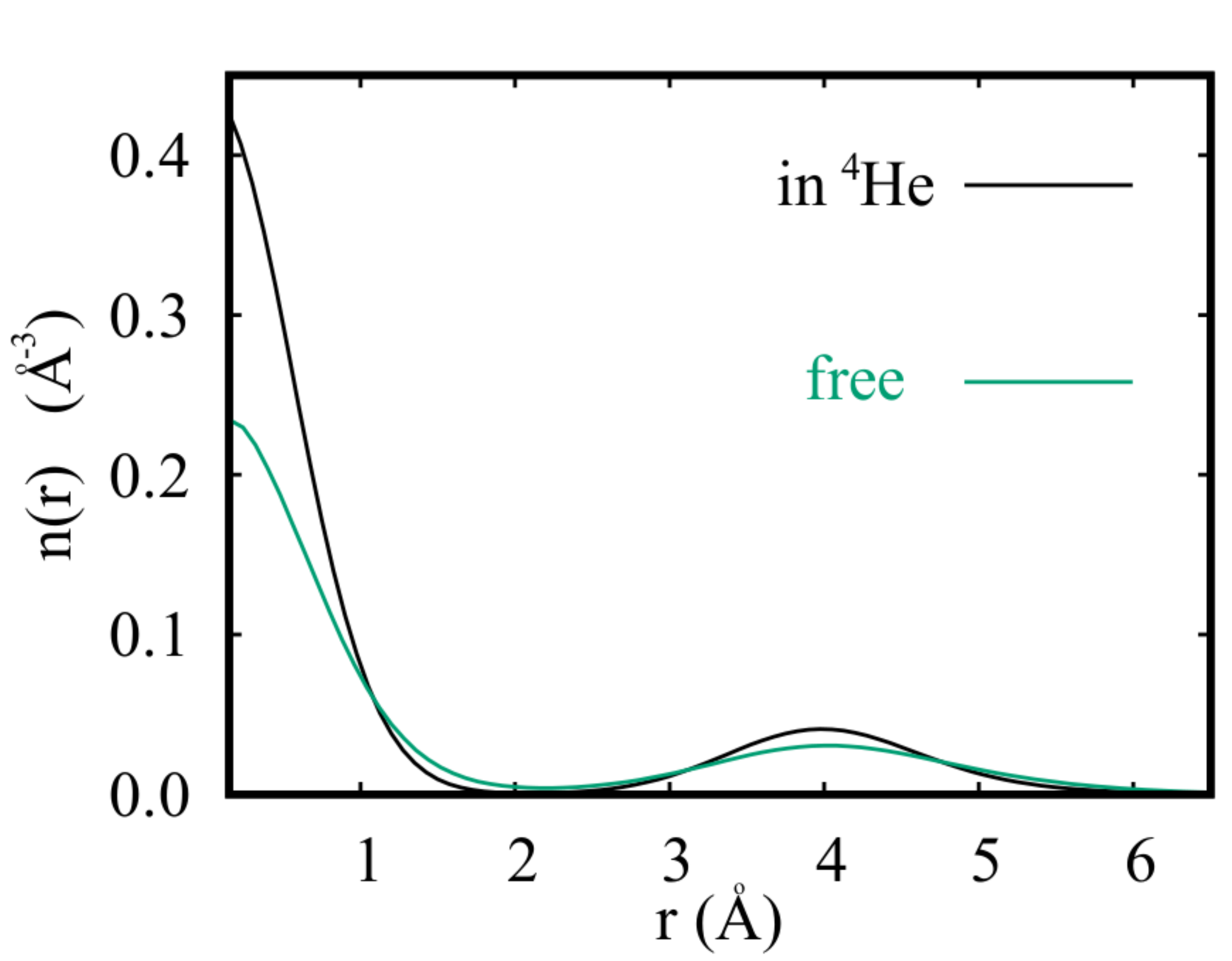}
\caption{{\rm Color online.} Radial density profile $n(r)$  (in \AA$^{-3}$) for a (\ph2)$_{15}$ cluster at temperature $T=0.25$ K, computed with respect to the center of mass of the cluster. Both the cases of a free cluster (i.e., in {\em vacuo}, lighter line), as well as in superfluid $^4$He (darker line) are shown.}
\label{radial}
\end{figure}
\\ \indent
We begin the illustration of our results with the observation that the (\ph2)$_{15}$ cluster does not break down but remains compact in superfluid $^4$He, and its structure is essentially temperature-independent below $T=1$ K. Fig. \ref{radial} shows the radial density profile, computed with respect to the center of mass, at a temperature $T=0.25$ K. Also shown for comparison is the corresponding profile {\em in vacuo}, at the same temperature. The structure is the same in both cases, with two well defined shells. However, \ph2 molecules in the cluster immersed in $^4$He feature a higher degree of localization, as shown by the greater height of the peaks, especially in the center of the cluster. The enhanced localization can be assessed quantitatively by looking at the kinetic energy per molecule, which is equal to $\sim$ 41 K, some 60\%  higher than that in the free cluster, worth just $\sim$ 25 K. 
Correspondingly, exchanges of \ph2 molecules, which are known to underlain superfluidity, are considerably less frequent than in the free cluster; for example, they are essentially non-existent $T=1$ K, a temperature at which the free cluster is 100\% superfluid, and remain exceedingly rare even at $T=0.5$ K (consistently, there is no evidence of a superfluid response of the cluster at these temperatures). Meanwhile, the surrounding  $^4$He matrix is superfluid at $T\lesssim 1$ K, the presence of the \ph2 cluster not affecting that to any significant degree \cite{noteb}.
\\ \indent
The superfluid fraction of the (\ph2)$_{15}$ cluster begins to be appreciably different from zero only at $T\lesssim 0.3$ K. For example, at $T=0.25$ K
it is equal to 0.44(4), while its value at $T=0.15$ K is 0.66(4).
\begin{figure}[ht]
\centering
\includegraphics[width=\linewidth]{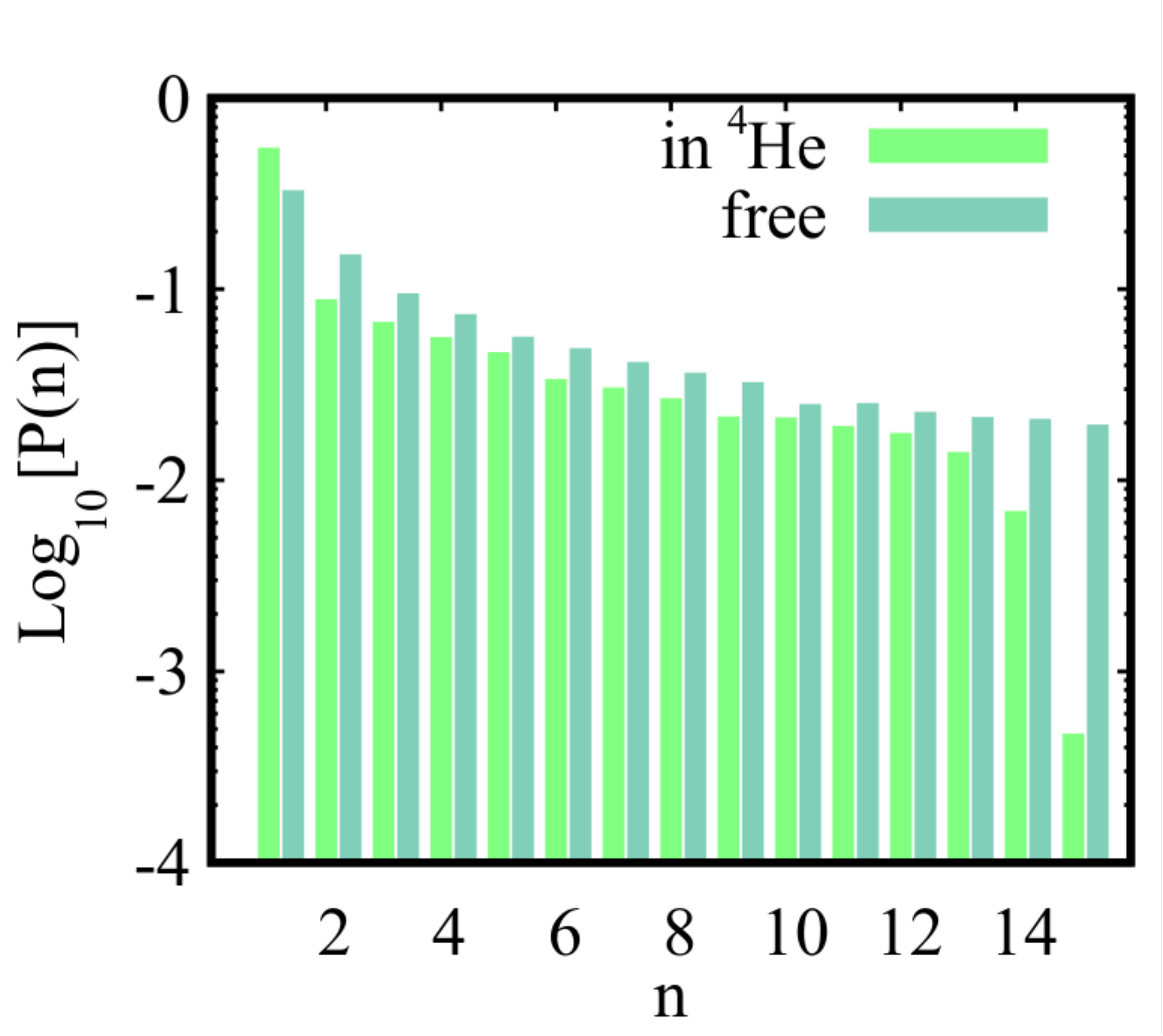}
\caption{{\rm Color online.} Logarithm (in base 10) of the frequency of occurrence of exchange cycles involving $n$ \ph2 molecules at $T=0.15$ K, for a free (\ph2)$_{15}$ cluster (lighter boxes) and for one immersed in $^4$He (lighter boxes). }
\label{cycle}
\end{figure}
\\ \indent
A quantitative assessment of the reduced propensity of a (\ph2)$_{15}$ cluster in superfluid $^4$He to display  quantum-mechanical exchanges, with respect to the same cluster {\em in vacuo}, can be obtained by comparing the computed frequency of occurrence $P(n)$ of exchange cycles involving $n$ \ph2 molecules for the two cases, shown in Fig. \ref {cycle}.
Although it does not directly correlate with the value of the superfluid response, $P(n)$ provides a direct measure  of the probability of occurrence of quantum mechanical exchanges in the cluster. 
In particular, exchanges involving (nearly) all of the molecules have been been shown \cite{fabiox,fabioy} to be crucial to the onset of superfluidity of a (\ph2) cluster. 
Fig. \ref{cycle} shows that, for a cluster immersed in superfluid $^4$He, the reduction is especially significant precisely for cycles involving {all} of the molecules in the cluster. At the same time, long  exchanges clearly do occur  at this low temperature, even if the cluster immersed in superfluid $^4$He, consistently with the presence of a finite superfluid response.  
\\ \indent 
We now discuss the relevance of the results of this first principle simulation of a (\ph2)$_{15}$ cluster in superfluid $^4$He at low temperatures to experiments whose interpretation is still under debate.
The results obtained in this work seem to lend  support to the contention that the (\ph2)$_{15}$ cluster immersed in superfluid $^4$He should turn superfluid at a temperature consistent with that of Ref. \onlinecite {grebenev}. The question, of course, is the degree to which the experimental conditions of Ref. \onlinecite{grebenev}, chiefly the presence of a dopant molecule (CO$_2$), might quantitatively and qualitatively alter the physical behavior of the system, as predicted theoretically in this work. Theoretical calculations on free clusters \cite{fabiox} have shown that the mere replacement of one \ph2 molecule with a heavier \od2, whose interaction with \ph2 molecules is very nearly the same, has a disruptive effect on the superfluid response of a cluster such as (\ph2)$_{15}$. One could certainly expect a heavier and more complex object, such as a (CO$_2$), to have at least equally significant an effect; however, theoretical evidence has been reported of superfluidity at temperature as high as 0.5 K  for a free  (\ph2)$_{15}$ cluster with an embedded CO$_2$ \cite{li}. This suggests that the main physical agent causing the suppression of superfluidity in small clusters immersed in superfluid $^4$He nanodroplets and with embedded linear molecules, is the surrounding $^4$He atoms, and that the effect of the embedded linear molecule may be quantitatively less important. The results presented here indicate that superfluidity is recovered if the temperature is lowered to $\sim 0.2 $ K, 
and therefore the interpretation of the experimental results of Ref. \onlinecite {grebenev} is plausible.
\\ \indent 
As a general remark, experiments  with molecules embedded in a superfluid $^4$He matrix have long been predicated on the tenet that superfluid helium should act like ``vacuum" \cite{cep}. Indeed, this calculation shows that $^4$He matrix leaves the cluster structure essentially unchanged, with respect to the that of the free cluster; however, the interactions of the $^4$He atoms with the \ph2 molecules, albeit relatively weak, have the effect the effect of rendering the \ph2 molecules less mobile and therefore suppressing quantum exchanges, consequently shifting the onset of superfluid behavior downward in temperature by an order of magnitude.
\\ \indent
This work was supported in part by the Natural Sciences and Engineering Research Council of Canada (NSERC). Computing support of Compute Canada is also acknowledged.


\begin{thebibliography}{99}
\bibitem{notex}
The results for free clusters discussed here were obtained in Ref. \onlinecite{fabio2}.
\bibitem{noteb}
In a system with ${\cal N}=256$ particles, a small reduction ($\sim$ 10\%) of the superfluid fraction of $^4$He arises from the breaking of translational invariance caused by the presence of the \ph2 cluster. However, the cycle length distribution is very close to that of bulk superfluid $^4$He. On a system comprising ${\cal N}=1024$ particles (i.e., 1009 $^4$He atoms), the superfluid fraction of $^4$He at $T=1$ K is essentially 100\%.
\end{thebibliography}
\end{document}